\documentstyle[aps]{revtex}
\input epsf
\begin{document}
\draft
\title{Synchronization in the non-extensive Kuramoto model}
\author{M. Bahiana}
\address{Instituto de F\'\i sica, UFRJ, Caixa Postal 68528, Rio de Janeiro, RJ, Brazil, 21945-970}
\author{M.S.O. Massunaga}
\address{Laborat\'orio de Ci\^encias F\'\i sicas, Universidade
Estadual do Norte Fluminense, Av. Alberto Lamego 2000}
\date{\today}
\maketitle
\begin{abstract}
We study the onset of synchronization in square lattices of limit cycle oscillators with long-range coupling by means of numerical simulations of the Kuramoto model. In this regime the critical coupling strength depends on the system size and interaction range reflecting the non extensive behavior of the system, but an adequate scaling removes the dependency and collapses the long-range synchronization curves with the one resulting from a system with uniform coupling. 
\end{abstract}
Coupled limit-cycle oscillators have been extensively used to model the collective behavior of systems in physics \cite{fisher}, chemistry and biological sciences \cite{glass-book1,win-book1}.
One of the simplest models to describe  those systems is the Kuramoto model for the time evolution for the phases of the oscillators. As shown in \cite{kura-book}, it possible to reduce any system of $N$ weakly coupled limit cycle oscillators near the Hopf bifurcation, to the set of equations
\begin{equation}
\dot{\theta}_i=\omega_i+\sum_{j}^{N}\Gamma_{ij}(\theta_j-\theta_i),\;\;\;(i=1,2,\ldots,N)\label{kmodel}
\end{equation}
where $\theta_i$ and $\omega_i$ are the phase and natural frequency of the $ith$ oscillator respectively. The coupling term $\Gamma_{ij}(\theta_i-\theta_j)$ is a periodic function that may be obtained from the original equations of motion, regardless of the dimensionality of the oscillating field. The case of uniform, or mean field,  coupling has been considered with detail by Kuramoto \cite{kura-book} for the simplest form of the coupling term,
\begin{equation}
\Gamma_{ij}=N^{-1}K\sin(\theta_j-\theta_i), \;\;\; (K>0)\label{coupling}
\end{equation}
where the factor $N^{-1}$ appears in order to ensure that the typical strength of the net local field experienced by each oscillator is independent of the total number $N$. Kuramoto considered the possibility of a collective oscillatory motion when the natural frequencies are picked from a given distribution $g(\omega)$, as would happen in a real system. As the coupling modifies the frequencies, acting as an external oscillatory force, it is necessary to define asymptotic frequency of the $ith$ oscillator after the transient regime as
\begin{equation}
\tilde{\omega}_i=\lim_{T\rightarrow\infty}\frac{1}{T}\left[\theta_i(t+T)-\theta_i(t)\right].\label{afreq}
\end{equation}
The existence of a collective behavior with phase and frequency entrainment depends on the coupling strength and on the choice of $g(\omega)$ \cite{kura-book,dai1}. In what follows we consider only the case where $g(\omega)$ is Gaussian.
 Kuramoto characterized the degree of synchronization in the asymptotic regime  by means of two order parameter like quantities: $R$ for the frequencies, and $\sigma$ for the phases. Frequency entrainment may be quantified simply by analyzing the formation of frequency clusters  and defining
\begin{equation}
R=\lim_{N\rightarrow \infty}\frac{N_s}{N},
\end{equation}
where $N_s$ is the size of the largest cluster of oscillators with a common asymptotic frequency $\tilde{\omega}$. 
The complex quantity
\begin{equation}
\sigma =\frac{1}{N}\sum^N_{j=1}\exp(i\theta_j)
\end{equation}
is appropriate for the phase ordering as it measures the concentration of phases on a certain value.
$R$ and $\sigma$ behave as  equilibrium order parameters in the sense that, for $K$ below some critical value $K_c$, oscillations are incoherent leading to $R$ and $\sigma$=0. For strong enough coupling, or $K>K_c$, collective oscillation takes place and $R,\sigma>0$. 

For the case of uniform coupling, phase and frequency entrainment appear simultaneously at the critical value
\begin{equation}
K_c=\frac{2}{\pi g(\omega_0)},\label{kc}
\end{equation}
where $\omega_0$ is the average value of the distribution of natural frequencies \cite{kura-book}.

The synchronization in lattices of oscillators with nearest neighbor coupling has also been studied with equilibrium statistical mechanics tools. The existence of a critical dimension for synchronization was questioned in \cite{ssk1} as macroscopic synchronization would require $R$ to be finite in the limit $N\rightarrow\infty$. This issue was clarified analytically in \cite{dai1} and numerically in \cite{aoygi} and\cite{bamass1}. The behavior of the synchronization curve $R\times K$ for different systems sizes in two dimensional lattices clearly showed that there is a limiting curve corresponding to the thermodynamical limit \cite{aoygi,bamass1} and, although far from equilibrium, the system is extensive. As opposed to the mean field case, there is no need for correction factors in the coupling term.

 Here we focus on systems with long range distance dependent coupling of the form
\begin{equation}
\Gamma_{ij}=\frac{K}{r^\alpha_{ij}}\sin(\theta_j-\theta_i),\;\;\;K>0,\label{necoupling}
\end{equation}
where $r_{ij}$ is the distance between oscillators $i$ and $j$. The limit $\alpha=0$ corresponds to uniform coupling, and $\alpha=\infty$ to nearest neighbor coupling. 
 From the results obtained in those limits, one can expect that there is a value of $\alpha$ below which the system is non-extensive due to the long range interactions. In Hamiltonian systems the limit of non-extensivity can be found by requiring the upper value of the energy to be finite as $N\rightarrow \infty$. Limit cycle oscillators are open systems and energy is a meaningless quantity so, instead, we establish the finiteness condition for $(1/N)\sum_i\dot{\theta}_i$ in the asymptotic regime. This condition may be established by requiring that the upper value of the coupling term divided by $N$ must be finite, as in the mean field coupling case.
With this, we end up with a condition identical to the one found for extensivity of the Ising model with distance dependent coupling \cite{ante-tsallis,cannas-tamarit}, for $\alpha< d$ the system is non-extensive. 
For any value of $d$ and $\alpha$, it is possible to incorporate the dependence on $\alpha$ and $N$ of the coupling term into the value of $K$  with the definition of a scaled coupling strength
\begin{equation}
K^*(\alpha,N)=K N^*,  \label{scaling}
\end{equation}
where 
\begin{equation}
N^*(\alpha,N)=\int ^{N^{1/d}}_1 dr r^{d-1}r^{-\alpha}=
\frac{N^{1-\alpha/d}-1}{1-\alpha/d}
\end{equation}
For $\alpha>d$ (\ref{scaling}) provides a correction for finite size effects, and for $\alpha<d$ it removes the dependency of the critical coupling strength on $\alpha$ and $N$. For the mean-field coupling ($\alpha=0$) and $N^*=N$, for $\alpha=d$ $N^*$ behaves like $\ln N$.

In order to numerically study the non-extensive region, we used two dimensional square lattices of coupled oscillators with natural frequencies sorted from the Gaussian distribution  $g(\omega)=(1/\sqrt{2\pi})\exp(-\omega^2/2)$. Equation (\ref{kmodel}) with the coupling (\ref{necoupling}) was then integrated using the Euler method with d$t=0.1$. The first 3000 iterations where discarded as a transient, and after 6000 iterations past the transient, we calculated the asymptotic frequencies as defined in (\ref{afreq}). We used the Hoshen-Kopelman method \cite{cluster} to analyze the frequency pattern, two sites were considered to belong to same cluster if the frequency difference between them was less than $2\pi/T$, where $T$ is the time elapsed after the transient. $R$ was then associated to the largest cluster, and averaged over 10 initial conditions for the natural frequencies. 

\begin{figure}[htb]
\leavevmode
\centering
\epsfxsize = 6cm
\epsfysize = 12cm
\epsffile{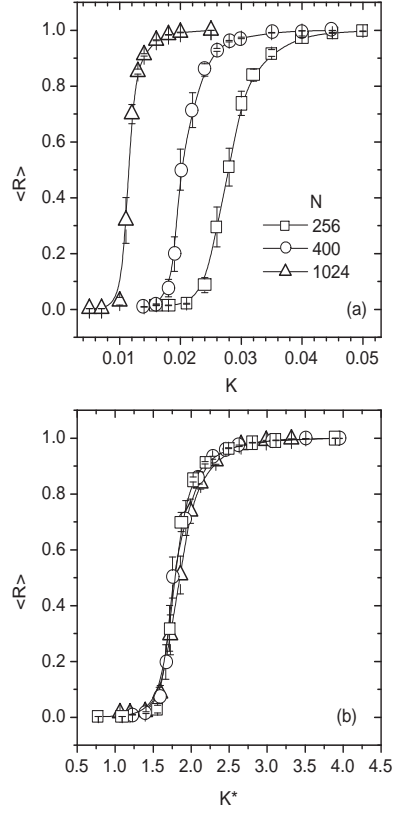}
\centering
\caption{Frequency entrainment in square lattices with $N$ oscillators interacting through a long range coupling $K/r^{0.75}$. (a) The non-extensivity appears in the $N$ dependency of $K_c$, the critical value for synchronization. (b) The corrected coupling intensity, $K^*$, absorbs the non-extensive part of the coupling term leading to a synchronization curves independent of the system size.}
\label{alfa075}
\end{figure}

The non-extensivity is evident if we consider the behavior of $\langle R\rangle$ as a function of $K$ for $\alpha<d$. For fixed $\alpha$, the critical  value of $K_c$  decreases with increasing $N$ as can be seen in Fig. (\ref{alfa075}.a), as more terms contribute to the coupling. The shift in $K_c$ may be corrected with the use of (\ref{scaling}).
In fact, if we plot $\langle R\rangle$ as a function of $K^*$, there is no dependence on the system size, the curves collapse and a unique value of $K_c^*$ appears. Figure (\ref{alfa075}.b) shows the behavior of $\langle R\rangle$ as a function of $K^*$ for fixed $\alpha$.    

\begin{figure}[htb]
\leavevmode
\centering
\epsfxsize = 6cm
\epsfysize = 12cm
\epsffile{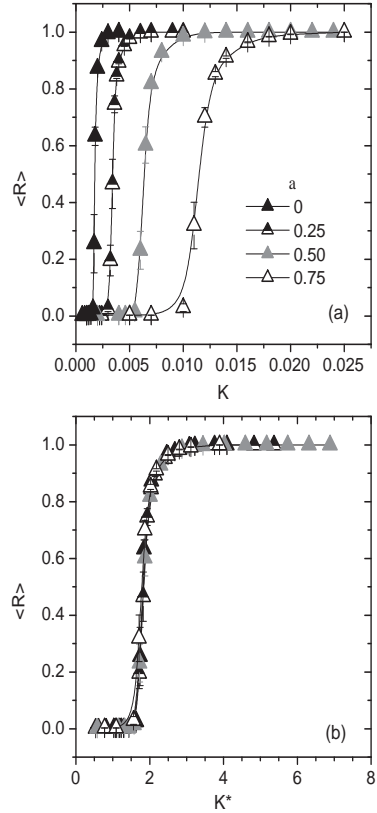}
\centering
\caption{Frequency entrainment in square lattices with 1024 oscillators interacting through a long range coupling $K/r^{\alpha}$. (a) The non-extensivity appears in the $\alpha$ dependency of $K_c$, the critical value for synchronization. (b) The corrected coupling intensity, $K^*$, absorbs the non-extensive part of the coupling term leading to synchronization curves independent of $\alpha$ The curve for $\alpha=0$ correspond to uniform or mean-field coupling.}
\label{N1024}
\end{figure}

The dependence of $K_c$ on $\alpha$ is evident if we plot $\langle R\rangle$ for fixed $N$ and different values of $\alpha$. Figure (\ref{N1024}.a) shows the curves for $\alpha=0.25$, 0.50, 0.75, and 0 (mean-field coupling without the correction factor $1/N$). Consistently with dependence on system size, as $\alpha$ decreases, more terms are included in the coupling, leading to smaller values of $K_c$. Again, (\ref{scaling}) fixes the dependence on $\alpha$ providing a perfect collapse even with the mean-field curve as can be seen in Figure (\ref{N1024}.b). Notice that the value of $K^*_c$ in Figures (\ref{alfa075}) and (\ref{N1024}) is in good agreement with (\ref{kc}), our frequency distribution leads to $K^*_c= 1.59$.  

In summary, we have extended the recently proposed equilibrium non-extensive scaling to a system of coupled limit-cycle oscillators far from equilibrium, by requiring the finiteness of the long-range coupling term $K/r^\alpha$. With this requirement it was possible to define a universal coupling strength $K^*$, independent of $\alpha$ and system size, allowing the collapse of the curves $\langle R\rangle\times K^*$ for different sizes and exponents, including the case of uniform coupling. This result allows the study of more general coupling forms, for example to include lattice deformation effects, for any range of the coupling term.

%
%

\end{document}